\documentclass[12pt]{iopart}

\usepackage{graphicx}
\usepackage[dvips]{epsfig}
\usepackage{iopams}  
\begin{document}

\title{Nature of resonant Raman scattering in ZnCoO films under sub-band excitation}

\author{K.A. Avramenko, V.P. Bryksa, V.V. Strelchuk,}

\address{Institute of Semiconductor Physics of NASU, Pr. Nauki 41, 03028 Kiev, Ukraine}

\author{C. Deparis, C. Morhain,}

\address{Centre de Recherches sur l'H\'{e}t\'{e}ro\'{e}pitaxie et ses Applications, CNRS, F-06560 Valbonne, France}

\author{and P. Tronc}

\address{Centre National de la Recherche Scientifique, Ecole Superieure de Physique et de Chimie Industrielles de la Ville de Paris, 10 rue Vauquelin, 75005 Paris, France}
\ead{bryksa@isp.kiev.ua}

\begin{abstract}
Using the methods of scanning electron (SEM) and atom force microscopy (AFM) as well as photoluminescence (PL) and Raman micro-spectroscopy, we investigated $Zn_{1-x}Co_xO$ films ($x=5$ and $15\%$) grown by molecular beam epitaxy on sapphire substrates. It is found that the films have a nanocrystalline structure with the grain size decreased from $\sim150$ down to $28$ nm at changing the $Co$ concentration from $5$ to $15\%$. High-resolution SEM images of $Zn_{0.85}Co_{0.15}O$ film have been interpreted as inhomogeneous lateral distribution of $Co$ atoms.

Two broad emission bands observed in spectra of band-to-band PL are ascribed to emission from $Zn_{1-x}Co_xO$ nano-regions enriched and depleted with $Co$. In low-temperature PL spectra under sub-band excitation of $Zn_{1-x}Co_xO$ films, there observed are intra-center optical transitions due to $Co^{2+}$ center: $^2E(G)$, $^2A_1(G)$, $^2T_1(G)$, $^4T_1(P)$, $^2T_2(G)$ $\longrightarrow$ $^4A_2(F)$.
Offered is a new approach to interpretation of resonant enhancement observed in multi-phonon scattering by $LO$ phonons in $Zn_{1-x}Co_xO$ under sub-bandgap excitation combined with the extrinsic Fr\"{o}hlich interaction mediated via a localized exciton bound to the isoelectronic impurity in which the electron is strongly localized at the magnetic $Co^{2+}$ ion. This conclusion is confirmed by the dependence of Raman spectra on the quantum energy of exciting radiation. It assumes formation of intermediated sub-band electron excited states of isoelectron $Co$ dopant, they are also referred to as charge transfer processes.

\end{abstract}
\maketitle

\section{Introduction}
$ZnO$ nanostructures with impurities of $3d$-metals are promising materials for devices in opto- and nanoelectronics, spintronics as direct-band material with a high width of the forbidden gap ($3.37$ eV) and high bound energy of excitons at room temperature ($\sim60$ meV)~\cite{Thomas}.

As it is commonly considered, doping the $ZnO$ epitaxial films with transition metals Co or Mn is able to essentially change the local structure of crystalline lattice, which influences on electron and magnetic properties of material. However, up to date the electronic structure of $3d$-centers in $ZnO$ remains unstudied in detail. It was shown that magnetic properties of $Co$-doped $ZnO$ films depend on the method of growing~\cite{Chambers} and structural defects~\cite{Schwartz}. Also observed was ferromagnetic ordering in $Zn_{1-x}Co_xO$ ($x=0.05-0.25$) with the Curie temperature $280-400$ $^\circ C$, and it was shown that growth in the charge carrier concentration provides enhanced ferromagnetism~\cite{Ueda}. For this case, exchange interaction between magnetic impurities is indirect and realized either via charge carriers~\cite{Dietl}, or defects and/or impurity electron states with a non-zero spin~\cite{Coey}, the magnetization value being essentially dependent on the concentration of free charge carriers and Fermi level position. The $n$-type conductivity in $ZnO$ is caused by oxygen vacancies $V_O$ ($E_d=0.3$ $eV$) as well as interstitial $Zn_i$ ($E_d=0.25$ $eV$)~\cite{Kohan}. Therefore, for $n$-type $Zn_{1-x}Co_xO$ where one can observe carrier-induced ferromagnetism above room temperature~\cite{Coey} with exchange interaction between localized spins of $d$-electrons in the $Co^{2+}$ center and electron states of shallow donors with a non-zero spin, offered is the so-called magnetic-polaron model of high-temperature magnetism. However, the energy of magnetic polaron is sufficiently low~\cite{Dietl1}, and, respectively, no quantitative experimental proofs to support the polaron model are found yet. There exist some other theoretical models where ferromagnetic ordering is realized due to inhomogeneous distribution of magnetic impurities in crystalline lattice~\cite{Sato}. Formation of a special network of nano-sized clusters with various $3d$-component composition (spinodal component decomposition) can be responsible for ferromagnetic ordering at temperatures higher than the room one~\cite{Sato}.

A non-significant difference between ionic radii of $Co^{2+}$ and $Zn^{2+}$ in tetrahedral coordination (ionic radius of $Co^{2+}$ is equal to $0.56$ $\AA{}$, $Zn^{2+}$ - $0.58$ $\AA{}$) should promote efficient dissolution of $Co$ in $ZnO$ and cause a weak distortion in crystalline lattice. As recently shown, $Co$ solubility in the wurtzite $ZnO$ matrix sharply increases with lowering the grain sizes of $ZnO$ crystallites. There obtained were MBE-grown $Zn_{1-x}Co_xO$ films with $x\leq15\%$ and sizes of crystalline grains $20-100$ $nm$~\cite{Straumal}. Also synthesized were colloidal $Zn_{1-x}Co_xO$ wurtzite nanocrystals ($d<10$ $nm$) with $0\leq x \leq1$~\cite{White}. Observed in solid solutions of wurtzite-type $Zn_{1-x}Co_xO$ was the effect of spinodal component decomposition and creation of local nano-regions enriched with magnetic impurities~\cite{Sato,Dietl2}. Similar nano-regions with spinodal component decomposition were found in DMS nano-structures ($d\sim50$ $nm$)~\cite{Kuroda}. It was shown that the local nano-regions of spinodal decomposition in the case of $Ga_{1-x}Mn_xAs$ DMS nano-structures are considerably less ($d\sim3$ $nm$)~\cite{Yokoyama}.

The method of Raman scattering is widely used for studying the phonon properties and secondary impurity phases in $Zn_{1-x}Co_xO$~\cite{Samanta}. The electron structure of defect-ion $Co^{2+}$ centers in $ZnO$ matrix was investigated using the method of resonant Raman scattering in the only work~\cite{Hasuike}. In this work, the resonant Raman process under sub-bandgap excitation of $Zn_{1-x}Co_xO$ polycrystals was explained as based on interaction of $LO$ phonons with carriers in excited states of $3d$-impurity center~\cite{Hasuike}. In $ZnO$ single crystals implanted with hydrogen and nitrogen atoms, the resonant multi-phonon scattering by n$LO$-phonons ($\omega_{LO}=577$ $cm^{-1}$) under sub-bandgap excitation ($2.71$ $eV$) was interpreted being based on the model for the Fr\"{o}hlich mechanism of interaction between $LO$-phonons and excitons bound with impurities~\cite{Friedrich}. On the other hand, as it was shown for $EuX$ ($X=S,Se,Te$) semiconductors, magnetic ordering changes the electron band structure, which can be pronounced in resonance effects observed in Raman spectra~\cite{Guntherodt}. As a result, one can observe strong multi-phonon processes in $LO(\Gamma)$-scattering that are the result of strong electron-phonon interaction related with excitation of $4f$-shell in $Eu$. In Raman spectra of $EuТе$, the lines of scattering in the vicinity $\omega_0=\omega_{LO}(k_L)$ and $2\omega_0$ are observed, and these are slowly shifted to $\omega_{LO}(k=0)$ and $2\omega_{LO}(k=0)$ with increasing the applied magnetic field. This resonant amplification of Raman scattering by $LO(\Gamma)$-phonons with end wave vectors in europium chalcogenides being in ferromagnetic and paramagnetic phases was interpreted with account of Fr\"{o}elich electron-phonon interaction and availability of spin fluctuations. Up to date, no investigation of resonant Raman scattering in DMS $Zn_{1-x}Co_xO$ under excitation with energies close to electron transitions with participation of $3d$-states was performed.

In this work, we used the complex of analytical methods of AFM and SEM microscopy as well as photoluminescence and Raman micro-spectroscopy to study electron and structure properties of MBE grown $Co$-doped $ZnO$ films. Especial attention was paid to studying the resonant Raman scattering caused by $Co^{2+}$ centers in $Zn_{1-x}Co_xO$ films with a nano-column structure. It has been shown that detailed analysis of Raman resonance effects in $ZnO$ films doped with $Co$ under sub-bandgap excitation can give information upon electron states of $Co^{2+}$ centers and the width of bandgap. Also, we have ascertained a clear dependence between the value (force) of electron-phonon interaction and correlation length that corresponds to diameters of the nano-column structure in $ZnO$ films, which are obtained using AFM and SEM structural-morphologic investigations.

\section{Experimental details} 
The $Zn_{1-x}Co_xO$ films were grown on $c$-sapphire substrates in a Riber Epineat MBE system equipment with conventional effusion cells for elemental $Zn$ and $Co$. Atomic oxygen was supplied via an Addon radiofrequency plasma cell equipped with a high-purity quartz cavity. The film thickness was about $1.7$ $\mu m$. X-ray diffraction patterns were used to ascertain the overall structure and phase purity. All positions of the peaks can be readily indexed to the hexagonal wurtzite $ZnO$ with lattice constants $a=3.2495$ $\AA{}$ and $c=5.2069$ $\AA{}$ for the undoped $ZnO$ sample, which matches well with the values of $ZnO$ single crystal. After $Co$ substitution with $Zn$ atom, both $a$- and $c$-axis lattice constants of $ZnO$ matrix were increased ($a=3.266$ ($3.259$) $\AA{}$ and $c=5.197$ ($5.195$) $\AA{}$ for $5$ ($15$) at.$\%$ $Co$, respectively). In particular, the ($002$) peak located at about $34.4^\circ$ is much stronger than the others for all the studied samples, which means that $ZnO$ films possess a nano-column structure with preferential orientation along $c$-axis.

Confocal micro-Raman and PL spectra collected by Jobin-Yvon T64000 triple spectrometer, equipped with a CCD detector, were performed using different lines of mixed $Ar^+/Kr^+$ ion laser ($457.9$ and $488.0$ $nm$) and $He$-$Cd$ laser ($325$ nm) with output power less than $10$ $mW$. An Olympus BX41 microscope equipped with a $\times100$ objective possessing $NA=0.90$ at room temperature measurements and a $\times50$ long-focus objective with $NA=0.60$ at low temperatures was used to focus laser light on the sample and collect scattered light into the spectrometer. Spatial resolution (lateral and axial) was about $1$ $\mu m$. The temperature dependent micro-Raman and PL spectra were performed using a Linkam THM600 temperature stage and CRYO Industries RC102-CFM helium cryostat. The AFM measurements were performed by a Dimension 3000 Nano-Scope IIIa scanning probe microscope. The chip structure of $Zn_{1-x}Co_xO$ samples were also studied using a ZEISS EVO-50 scanning electron microscope (SEM).

\section{Results and discussion} 
\begin{figure}
\centerline{\includegraphics[angle=0,width=0.6\textwidth]{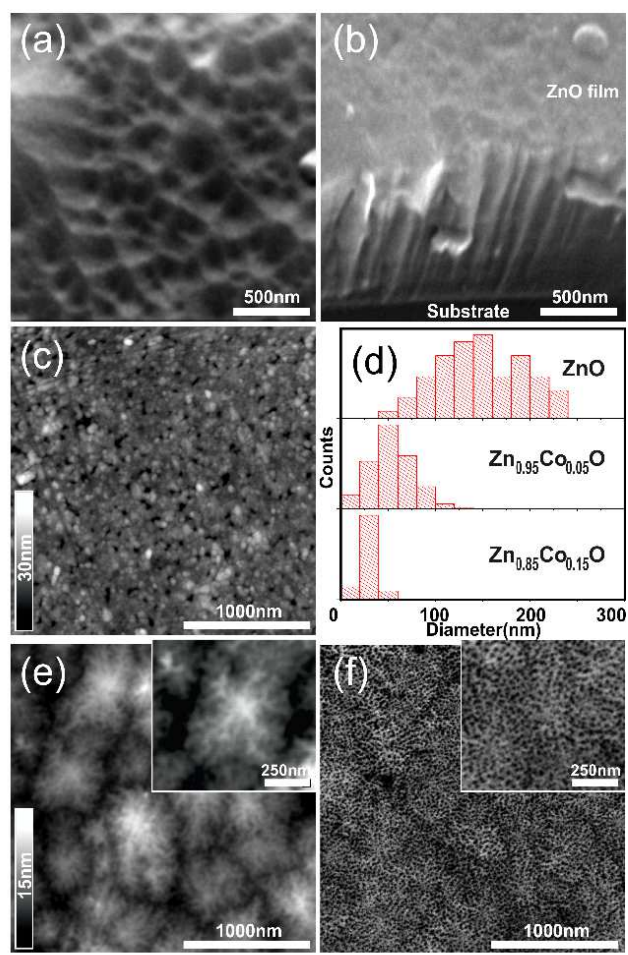}}
\bigskip
\caption{\label{fig01} SEM images of the surface (a) and cross section (b) of the non-doped film and AFM images of surfaces of $ZnO$ films doped with $5\%Co$ (c) and $15\%Co$ (e). High-resolution SEM images of $ZnO$ film doped with $15\%Co$ (f). Shown in the insert (e, f) is a separate fragment of $15\%Co$ $ZnO$ film obtained using AFM and high-resolution SEM methods. Statistically averaged grain size distribution for the ZnO film non-doped (d), doped with $5\%Co$ (c) and $15\%Co$ (e).}
\end{figure}

Shown in Fig.~\ref{fig01} are images of surface with its typical morphology for three studied MBE-grown $ZnO$ epitaxial films and their corresponding cross-section of the SEM image. These $ZnO$ films possess a nanograin structure of surface morphology with the mean grain size $\sim140$ $nm$ (Figs~\ref{fig01}а, \ref{fig01}d).

The SEM images near the chipped edge of the studied $ZnO$ film show a column-like structure along the growth direction (Fig.~\ref{fig01}b). In the case of $5\%Co$ $ZnO$ (Fig.~\ref{fig01}c), the main grain size is  decreased approximately down to $\sim40$ $nm$ (Fig.~\ref{fig01}b) as compared to the undoped sample. In the case of $15\%Co$ $ZnO$ film, a complex structure consisting of large blocks (with the size $\sim500$ $nm$) (Fig.~\ref{fig01}e), everyone of which possesses a substructure with nano-grains of mean sizes close to $28$ $nm$ (Figs~\ref{fig01}d and \ref{fig01}f). In the case of $15\%Co$ $ZnO$ films, we observed a difference between AFM surface morphology and SEM image obtained with high spatial resolution approximately $2$ $nm$. It is related with the fact that in the SEM image, beside surface relief, considerable contribution is provided by the phase contrast caused by inhomogeneous distribution of $Co$ atoms in $ZnO$ film in the sub-surface of the film with the thickness up to $\sim2$-$3$ $nm$. The compositional phase contrast is caused by the considerably higher efficiency of electron absorption in $Co$ atoms as compared with those of $ZnO$ matrix ($2$- to $3$-times more). Therefore, it is reasonable to suppose that the observed clear contrast in SEM images in the form of light and dark regions (Fig.~\ref{fig01}f) may be caused by inhomogeneities in lateral distribution of $Co$ atoms for $15\%Co$ $ZnO$ sample in the nanometer scale level. This effect can serve as a direct experimental confirmation for formation of enriched and depleted with $Co$ local regions in $ZnO$ matrix. It seems interesting to note that in $15\%Co$ $ZnO$ films mean sizes of light regions in Fig.~\ref{fig01}f (i.e., $Co$-enriched nano-regions) are equal to approximately $100$ $nm$ as compared with dark regions corresponding to lower $Co$ concentrations ($\sim28$ $nm$). A similar effect was observed in MBE grown $ZnTe$ films doped with $Cr$ atoms in SEM measurements with nanometer spatial resolution~\cite{Kuroda}. The authors interpreted their obtained maps as inhomogeneous lateral distribution of $Cr$ atoms in local regions of the size close to $50$ $nm$ as spinodal component decomposition of the $Zn_{1-x}Cr_xSe$ alloy, which in accord with the theoretical investigation~\cite{Sato} can cause appearance of ferromagnetism at room Curie temperatures.

Performed X-ray diffraction investigation of $ZnO$ films doped with $Co$ did not found any secondary structural phase in them within the detection limit. The observed inhomogeneous distribution of the doping $Co$ impurity is possibly caused by MBE growing process realized in conditions far from the equilibrium ones, when there formed are strained $Zn_{1-x}Co_xO$ films with a grain structure related with spinodal component decomposition. It is known that this MBE growth allows essentially higher concentrations of magnetic impurities than the equilibrium solubility limit~\cite{Straumal}.

\begin{figure}
\centerline{\includegraphics[angle=0,width=0.5\textwidth]{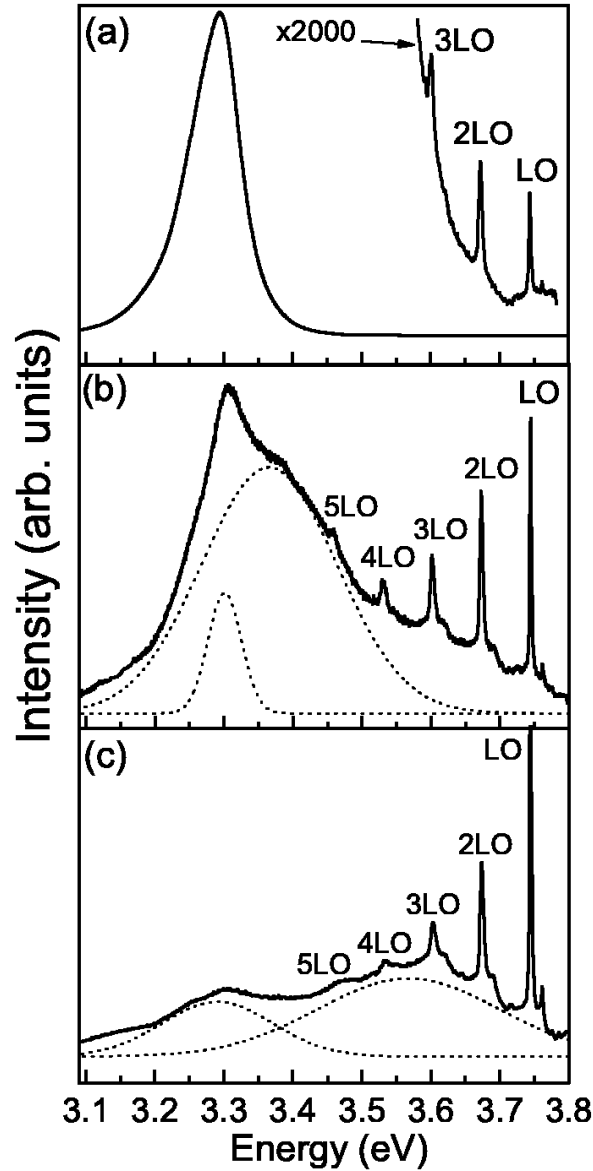}}
\bigskip
\caption{\label{fig02} PL spectra of non-doped (a), doped with $5\%Co$ (b) and $15\%Co$ (с) $ZnO$ epitaxial films. $T=300$ K, $E_{exc}=3.81$ eV. Dashed lines indicate Gaussian contours for decomposition of NBE emission.}
\end{figure}

\begin{figure}
\centerline{\includegraphics[angle=0,width=0.5\textwidth]{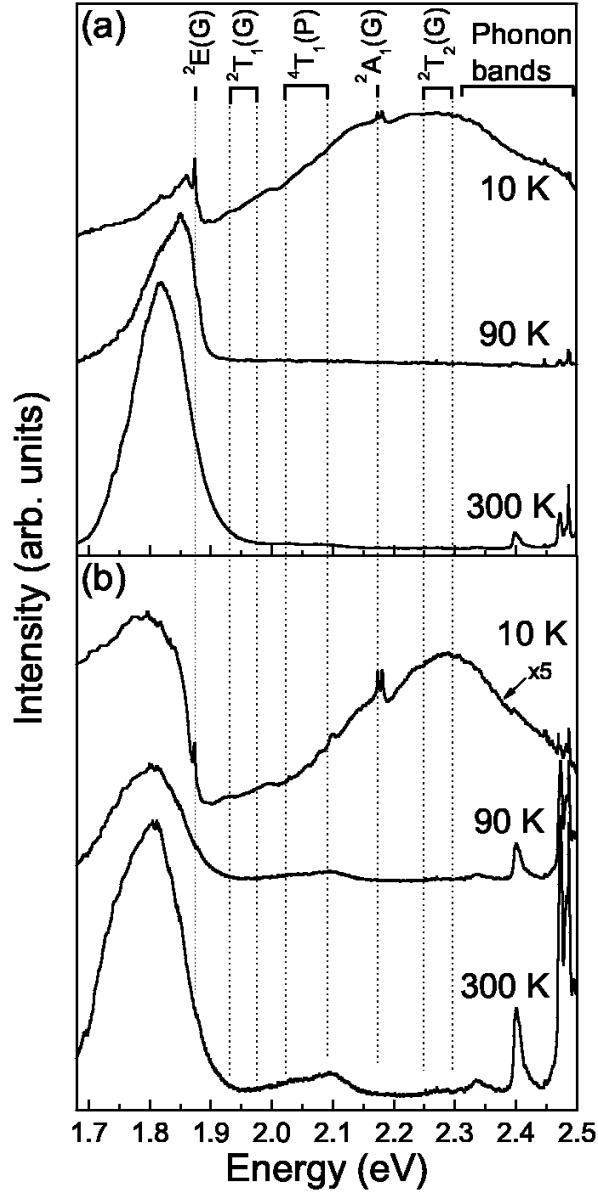}}
\bigskip
\caption{\label{fig03} PL spectra of $ZnO$ films doped with $5\%Co$ (a) and $15\%Co$ (b). $E_{exc}=2.54$ eV. Dotted lines indicate term energies for $Co^{2+}$ ions in tetrahedral crystalline surrounding. PL bands of the $Cr^{3+}$:$Al_2O_3$ substrate are subtracted from the spectra.}
\end{figure}

Shown in Fig.~\ref{fig02}a is the spectrum of near-band-edge (NBE) emission of the non-doped $ZnO$ film. There one can see the intense band of NBE PL with the peak at $\approx3.3$ $eV$ and halfwidth $\approx70$ $meV$, which corresponds to emission of excitons bound at donor and acceptor states~\cite{Chen}. From the low-energy side, asymmetry in the PL band shape can be modeled with Gaussian lines with the distance between their peaks close to the $LO$ phonon energy ($\approx70$ $meV$), which can be indicative of a high efficiency of recombination processes related with excited charge carriers and $LO$ phonons.

In PL spectra of $ZnO$ films doped with $Co$ concentrations $5$ and $15\%$ (Fig.~\ref{fig02}b, с), one can observe a considerable decrease ($\sim1000$- and $5000$-times) in the intensity of NBE PL bands, respectively. In this case, the PL band peak is shifted to the high-energy side by $\Delta E\approx15$ $meV$, which can be indicative of a quantum-sized effect or $sp$-$d$  electron band mixing. As the dimensions of nanocrystalline regions ($d>30$ $nm$, Fig.~\ref{fig01}) are considerably higher than the exciton radius in ZnO ($\sim1$ $nm$)~\cite{Fonoberov}, the quantum-sized effect can be neglected. When the $Co$ concentration increases, there appears and increases the PL band in the high-energy spectral range. It can be modeled by wide Gaussian contours peaking at $\sim3.36$ $eV$ ($\Gamma\approx 63$ $meV$) and $\sim3.57$ $eV$ ($\Gamma\approx 189$ $meV$) for $5\%Co$ and $15\%Co$ samples, respectively (Figs~\ref{fig02}b and \ref{fig02}с). The latter are ascribed by us to emission from two separate regions in $Zn_{1-x}Co_xO$, namely: regions enriched and depleted with $Co$ atoms, which in a good accordance with SEM investigations (Fig.~\ref{fig01}). It should be noted that the observed high-energy shift of both NBE PL bands by $\Delta E\sim15$ and $200$ $meV$ cannot be explained by increasing of the $Co$ concentration in $Zn_{1-x}Co_xO$ alloy, as it is known that in the wurtzite structure $CoO$ $E_g\sim2.3$ $eV$~\cite{White}. This blue shift can be caused by the $sp$-$d$ mixing of wave functions inherent to electrons of the conduction band with localized $d$-states of magnetic impurity. This effect is observed in ferromagnetic semiconductors $Ga_{1-x}Mn_xAs$ ($\Delta E\approx150$ $meV$) and $EuO$ ($\Delta E\approx300$ $meV$) where the high-energy shift of the intrinsic absorption edge $\Delta E$ increases with magnetization~\cite{Lebedeva}.

Also, NBE PL spectra of $Zn_{1-x}Co_xO$ films show weak additional emission at $E>3.7$ $eV$ caused by relaxation of hot charge carriers with a high concentration ($\sim10^{19}$-$10^{20}$ $cm^{-3}$).

It is worth to note that the intensity of the broad band within the range $2.0$ to $2.3$ $eV$ related with defects~\cite{Zeng}, $V_O$~\cite{Kurbanov},\cite{Gong}, $O_i$~\cite{Zubiaga} is three orders lower in the non-doped $ZnO$ films than that of NBE PL band in $Zn_{1-x}Co_xO$ films. Depicted in Fig.~\ref{fig03} are the PL spectra of $Zn_{1-x}Co_xO$ films at various temperatures for the excitation energy lower than the edge of intrinsic absorption in $ZnO$ matrix. In $ZnO$ films doped with $Co$, one can observe the splitting of $d^7$ configurations inherent to $Co^{2+}$ ions by electron terms as a result of crystalline field effects in tetragonal surrounding of oxygen atoms. The parameter of crystalline field in $ZnO$ is $\Delta=5263$ $cm^{-1}$ and electron transitions between the ground $^4A_2(F)$ and excited $^2E(G)$, $^2T_1(G)$, $^2A_1(G)$, $^4T_1(P)$ terms of the $Co^{2+}$ ion lie in the forbidden gap of $ZnO$ matrix (Fig.~\ref{fig03}а)~\cite{Koidl}.

The narrow emission bands at $1.88$ and $2.18$ eV in low-temperature PL spectra of $5\%$ and $15\%Co$ samples ($T=10$ K, Fig.~\ref{fig03}) correspond to electron transitions $^2E(G)$ $\rightarrow$ $^4A_2(F)$ and $^2A_1(G)$ $\rightarrow$ $^4A_2(F)$ in the $Co^{2+}$ center, respectively. The broad emission bands at $~\sim1.97$, $2.05$ and $2.3$ eV correspond to electron transitions $^2T_1(G)$ $\rightarrow$ $^4A_2(F)$, $^4T_1(P)$ $\rightarrow$ $^4A_2(F)$ and $^2T_2(G)$ $\rightarrow$ $^4A_2(F)$, respectively (see the notations of terms for the $Co^{2+}$ center in Fig.~\ref{fig03}). At the background of the narrow band corresponding to the electron transition $^2E(G)$ $\rightarrow$ $^4A_2(F)$ in the $Co^{2+}$ center, one can observe a broad low-energy emission band that was earlier explained by participation of phonons localized in distorted crystalline field in this electron transition.
With increasing the temperature ($Т>50$ K), PL spectra show a considerable decrease in the intensity of emission bands within the range of high-energy electron transitions in the $Co^{2+}$ center, and sharp peaks disappear. However, the intensity of the PL band related with the lowest electron transition $^2E(G)$ $\rightarrow$ $^4A_2(F)$ increases in the temperature range from $10$ to $300$ K, and the band becomes widen and shifted to the low-energy side by $\Delta E\approx70$ $meV$ (Fig.~\ref{fig03}), which can be explained both by the influence of electro-phonon processes and broadening the ground level of the interacting $Co^{2+}$ centers.

As known, the electron terms $^2T_1(G)$, $^4T_1(P)$ and $^2T_2(G)$ take an active part in charge transfer processes for electrons from/onto the $Co^{2+}$ center with creation of $Co^{1+}$ and $Co^{3+}$ states~\cite{White}. Therefore, it seems reasonable to assume that the absence PL bands related with these terms at room temperature can be indicative of availability of nonemission electron transitions from these states to the conduction band of $ZnO$ matrix.

\begin{figure}
\centerline{\includegraphics[angle=0,width=0.5\textwidth]{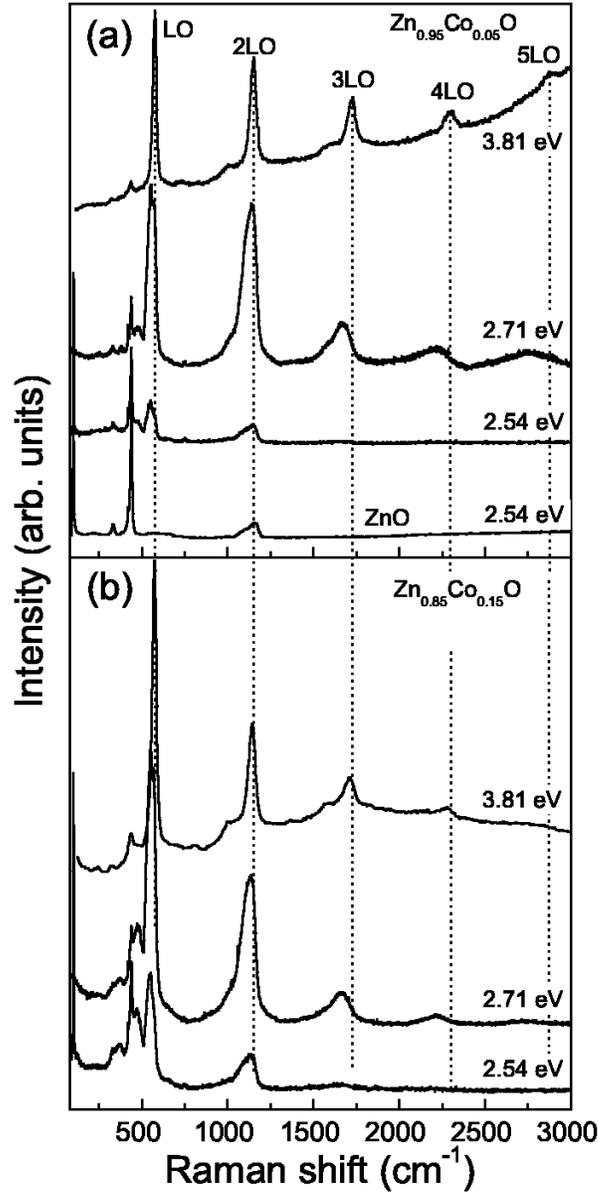}}
\bigskip
\caption{\label{fig04} Micro-Raman spectra of the non-doped $ZnO$ film (a) and $Zn_{1-x}Co_xO$ films for $x=5$ (a) and $15\%$ (b) obtained at various values of the excitation energy. $T=300$ K. The spectra have been normalized by the intensity of the $A_{1g}$ phonon mode of the sapphire substrate at $418$ $cm^{-1}$.}
\end{figure}

\begin{figure}
\centerline{\includegraphics[angle=0,width=0.5\textwidth]{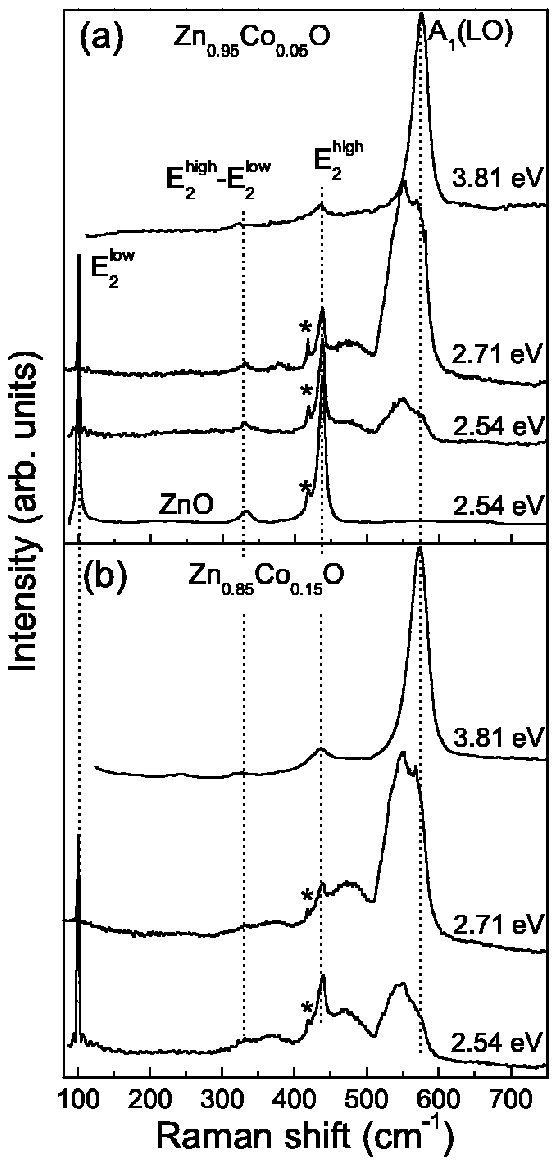}}
\bigskip
\caption{\label{fig05} Micro-Raman spectra of the first order within the range $80$ to $750$ $cm^{-1}$ for $Zn_{1-x}Co_xO$ films at $x=5$ (a) and $15\%$ (b) obtained at various values of the excitation energy. $T=300$ K. The asterisk indicates the $A_{1g}$ phonon mode of the sapphire substrate.}
\end{figure}

Phonon properties of $ZnO$ and $Zn_{1-x}Co_xO$ films were investigated using micro-Raman scattering in $Z(X,X)-Z$ geometry, where the axis $Z$ is directed along the wurtzite $c$-axis. $ZnO$ crystals possess the hexagonal wurtzite structure (spatial symmetry group $C^4_{6v}$). In accord with the theory-group analysis, in the $\Gamma$-point of the $ZnO$ Brillouin zone the phonon modes belong to the irreducible representation: $\Gamma=A_1(Z)+2B_1+E_1(X,Y)+2E_2$. The modes $A_1(Z)$, $E_1(X,Y)$ and $E_2$ are active in Raman spectra, while the modes of $B_1$ symmetry are the so-called ''silent'' modes, i.e., these cannot be observed in Raman spectra.
In non-resonant Raman spectra of non-doped $ZnO$ films with $E_{exc}=2.54$ eV, one can observe the intense  and  phonon modes at $\approx100.5$ $cm^{-1}$ ($\Gamma=1.6$ $cm^{-1}$) and $\approx439.5$ $cm^{-1}$ ($\Gamma=5.0$ $cm^{-1}$), respectively (Figs~\ref{fig04} and \ref{fig05}). The observed high-frequency shift of the mode by $\Delta\omega\approx2.5$ $cm^{-1}$ as compared with bulk $ZnO$ corresponds to bi-axial compression strains in the plane perpendicular to the $c$-axis. The difference two-phonon mode $E_2^{high}-E_2^{low}$ is observed at $\approx333$ $cm^{-1}$. The $A_1(LO)$ mode at $574$ $cm^{−1}$ possesses very low intensity in high-quality $ZnO$ films, which is related with destructive interference between Fr\"{o}ehlich interaction and deformation potential contribution to the $LO$ scattering in $ZnO$~\cite{Callender}. The observed phonon modes correspond to wurtzite $ZnO$ crystal~\cite{Cusco}. The asterisk in Fig.~\ref{fig05} indicates the intense phonon mode of $A_{1g}$ symmetry at $418$ $cm^{-1}$, which is inherent to sapphire substrate.

Doping the $ZnO$ films with $Co$ atoms causes structural disordering the crystalline lattice and stoichiometric changes in it. As the non-polar $E_2^{high}$ mode corresponds to oscillations of metal atoms, and $E_2^{low}$ - to oscillations of oxygen atoms in the plane normal to $c$-axis, they are very sensitive to disordering both in cation and anion sub-lattices, respectively. With increasing the $Co$ concentration in the studied samples, the intensity of these $E_2$ modes decreases. $E_2^{low}$ ($E_2^{high}$) mode is shifted to the red (blue) spectral ranges by $\sim0.7$ ($1.2$) $cm^{-1}$ and is broadened up to $\sim2.2$ ($\sim13.0$) $cm^{-1}$ as a consequence of structural disordering and violation of the translation symmetry in metal (oxygen) sub-lattices and/or creation of $Zn_{1-x}Co_xO$ solid solution.

Also, new additional broad bands appear within the range $550$-$600$ $cm^{-1}$ (Figs~\ref{fig04} and \ref{fig05})~\cite{Scepanovic} in Raman spectra of $Zn_{1-x}Co_xO$ films. In general, these bands can be caused by $Co$-impurity substitution in cation sublattice and correspond both to local resonant impurity oscillations and impurity-activated vibration bands related with forbidden Raman scattering with wave vectors of phonon density of states, in particular, as a consequence of selection rule violation~\cite{Zigone}. It is very difficult to distinguish between these two types of vibrations in Raman spectra, as their frequencies are close~\cite{Zigone}. It is appropriate to note that that forbidden Raman scattering was also observed in magnetic materials, in particular in halogenides of transition elements $EuX$ ($X=S,Se,Te$), where the multi-phonon $LO(\Gamma)$-process of scattering caused by spin disordering and mixing between electron and magnetic excitations is dominant~\cite{Guntherodt}. The broad band at $\sim471.0$ $cm^{-1}$ is ascribed to surface optical phonons (SP-mode) that can be detected with the frequency position between $TO$ and $LO$ modes in $ZnO$ nanostructures with the sizes $<100$ $nm$~\cite{Fonoberov1}. It should be noted that in Raman spectra of the studied $Zn_{1-x}Co_xO$ films, we did not found any additional phonon bands that could correspond to secondary structural phases such as $Co_3O_4$ or $Zn_xCo_{3-x}O_4$~\cite{Samanta}.

Shown in Fig.~\ref{fig04} are the Raman spectra of $Zn_{1-x}Co_xO$ films under subband excitation ($E_{exc}=2.54$ and $2.71$ $eV$) as well as under resonant excitation within the range of intrinsic absorption ($E_{exc}=3.81$ $eV$). In these spectra, one can see the bands caused by the multi-phonon $LO$-process of resonant Raman scattering that will be analyzed in detail below.

Fig.~\ref{fig05} illustrates well resolved Raman spectra of the first order for $Zn_{1-x}Co_xO$ films within the frequency range $80$ to $750$ $cm^{-1}$. It is known that structural defects and local fluctuations of the component composition for $Zn_{1-x}Co_xO$ alloy cause violation of the selection rules as to the wave vector ($\bf{q}=0$), which provides participation of phonons with arbitrary wave vectors in the process of Raman scattering. Thus, the sharp peaks in the intense band at $550\div600$ $cm^{-1}$ that is observed in various geometries of experiments can be related with the van Hove singularities or with critical points located in directions of high symmetry in the Brillouin zone and correspond to maximum values of the density of single-phonon states in ZnO~\cite{Serrano} and/or to the frequency of local Co-impurity resonant vibrations.

In spectra of resonant Raman scattering, when the energy of excitation quantum is close or higher than the energy of real electron states in semiconductor, one can detect bands that correspond to multi-phonon $LO$-process of scattering caused by the Fr\"{o}ehlich electron-phonon mechanism of interaction~\cite{Bergman}. It means that the resonant Raman scattering may be useful when studying the interband energy electron transitions in semiconductors. The multi-phonon scattering by $LO$-phonons was observed in resonant Raman scattering in monocrystalline bulk $ZnO$, $ZnO$ films and $ZnO$ nanowires. In all these cases, excitation of Raman spectra in samples was performed using the discrete emission line from a $He$-$Cd$ laser with the energy $E_{exc}=3.81$ $eV$ ($325$ $nm$) that is $\approx440$ $meV$ lower than the width of $ZnO$ forbidden gap, which corresponds to the condition of input resonance with interband electron transition. As seen from Fig.~\ref{fig04}, in the resonant Raman spectra of the studied $Zn_{1-x}Co_xO$ films, on the broad background of edge luminescent emission, one can observe $LO$-phonon modes at $574$ $cm^{-1}$, $1170$ $cm^{-1}$, ... for $5\%Co$ and at $574$ $cm^{-1}$, $1170$ $cm^{-1}$, ... for $15\%Co$ that correspond to $1LO$, $2LO$, $3LO$, $4LO$ and $5LO$ phonon bands of $Zn_{1-x}Co_xO$ (Figs~\ref{fig04} and \ref{fig05}). Rather extraordinary was the effect that when exciting the Raman spectra in $Zn_{1-x}Co_xO$ films with the quantum energy within the range $2.54$ to $2.71$ $eV$ (subband excitation), one can observe a considerable (approximately $10$-times) increase in the intensity of the $LO$ signal and multi-phonon $LO$-scattering up to the fifth order (Fig.~\ref{fig04}). It is unambiguous confirmation of the resonant Raman scattering process.

It is noteworthy that this resonant dependence was observed only in $Co$-doped $ZnO$ samples. Unusual was the observed low-frequency shift of all the $LO$-phonon repetitions under subband ($2.54$ and $2.71$ $eV$) as compared with band-to-band ($3.81$ $eV$) excitation in micro-Raman spectra obtained from the same local region of the sample. When using the subband excitation, it is rather difficult to estimate the value of frequency shifts with account of the frequency position inherent to $1LO$ band of the first and $2LO$ band of the second scattering order. This cumbersome situation can be caused by the fact that in our micro-Raman experiments we study a local region with the diameter $\sim500$ $nm$ where available are at least two nano-regions of $Zn_{1-x}Co_xO$ alloys with various component compositions (both enriched and depleted with $Co$, Fig.~\ref{fig01}). As a result, we have simultaneously observed two resonant processes of Raman scattering with different degree of efficiency in local regions of $Zn_{1-x}Co_xO$ samples. In the case of dominant resonant process in the $Zn_{1-x}Co_xO$ film, the frequency shift of $3LO$ line of the third order is close to $\sim65.6$ $cm^{-1}$ and $\sim48.9$ $cm^{-1}$ for concentrations $5$ and $15\%Co$. Note that the effect of frequency change for $LO$ phonons in resonant Raman spectra cannot be explained by only participation of the Brillouin zone with  in the multi-phonon process of phonon scattering, which was observed in $GaN$ films ($\omega_{2LO}<2\omega_{LO}$)~\cite{Kaschner}. In the opposite case, the $LO$-phonon frequency could be the same under subband and band-to-band excitation.

\begin{figure}
\centerline{\includegraphics[angle=0,width=0.5\textwidth]{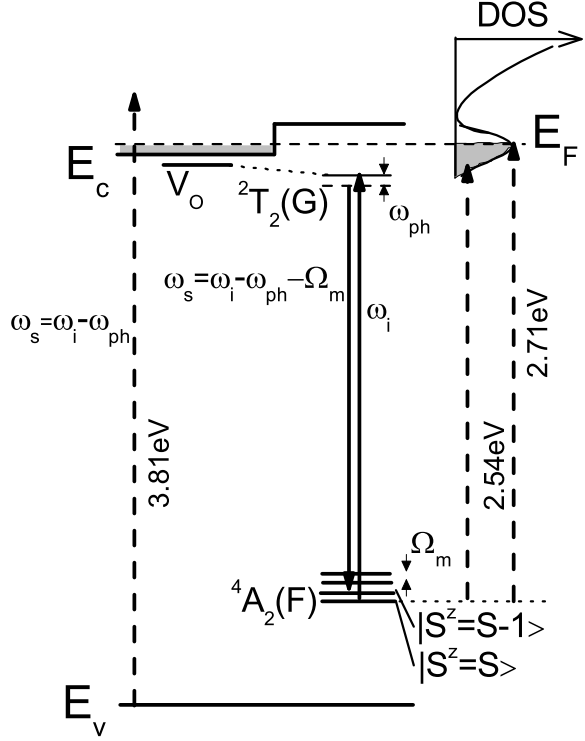}}
\bigskip
\caption{\label{fig06} Electronic states involved in the resonance Raman scattering for magnetically ordered phase of the $Co$ doped $ZnO$ films. The top of valence ($E_v$) and bottom of conduction ($E_c$) bands correspond to zero and $3.4$ $eV$, respectively. The Fermi level ($E_F$ ) indicates the degenerate semiconductor with the electron concentration $n\approx10^{20}$ $cm^{-3}$. It shows electronic transition of $Co^{2+}$ $3d^7$ orbitals by $^4A_2(F)$ and $^2T_2(G)$ terms. The localized magnetic moments of $Co^{2+}$ can long-range interact via the collective electrons of the conduction band and form the impurity sub-band near the bottom of conduction band. $\omega_i$($\omega_s$) is the frequency of the incident (scattered) phonon. $\omega_{ph}$($\Omega_m$) is the frequency of the phonon (magnon).}
\end{figure}

To explain the resonant character of Raman spectra under subband excitation, we have proposed the following energy diagram for $Zn_{1-x}Co_xO$ films (Fig.~\ref{fig06}). Using the determined from PL spectra (Fig.~\ref{fig03}) energy positions of emission bands that correspond to intra-center optical transitions from the ground state $^4A_2(F)$ to the excited states of electron transitions $^2T_1(G)$, $^4T_1(P)$ and $^2T_2(G)$ $\rightarrow$ $^4A_2(F)$ inherent to the ion $Co^{2+}$, we estimated the depth of the ground level for this center relatively to the edge of valence band as $0.8$ to $1.0$ $eV$. Due to partially filled $3d$ shells, the $Co$ impurity atoms can exist in various charge states. As to $ZnO$, it is adopted that change in the charge state of $3d$-impurity is possible either due to shallow donor centers (vacancies of oxygen, $V_O$, $E_d=25$ $meV$, interstitial zinc $Zn_i$, $E_d=25$ $meV$) or under photoionization by light. It was formed the bounding states due to the long-range Coulomb interaction as a result an electron occupies by the hydrogen-like orbital of the donors. The excited states of $3d$-defect centers together with the donor states of $ZnO$ host are formed a defect exciton~\cite{Nielsen}. So, it was observed the resonant Raman processes and a selective enhancement in multi-phonon scattering by $LO$ phonons. Our results show that the intermediated states in multi-phonon scattering by $LO$ phonons in $Zn_{1-x}Co_xO$ under sub-bandgap excitation is a excited state of $3d$-defect $Co^{2+}$ centers bounded to a the localized exciton with the Fr\"{o}hlich mechanism.
This approach of the resonant enhancement is a similar one observed for $EuX$ ($X=S,Se,Te$) semiconductors in multi-phonon scattering by $LO$ phonons which is due to a strong interaction phonons with localized electrons in non-filled $4f$-shells~\cite{Guntherodt}.

Existence of magnetic excitations related to the ground state of $Co^{2+}$ center with the frequency $\Omega_m$ in the ferromagnetic phase ($T<T_c$) provides creation of a magnetic sub-band that consists of the states $|S^z=S>$, $|S^z=S-1>$, etc~\cite{Guntherodt1}. Mixing the magnon-phonon excitations results in the low-frequency shift and broadening the orders of $LO$ phonons in resonant Raman spectra under sub-band excitations in $Co$-doped $ZnO$ films. The analysis of low-frequency shifts of $LO$-phonon repetitions in the Raman spectra under resonant and non-resonant excitation conditions allows estimating the band width for magnetic sub-system.

\section{Conclusion}
In this paper, we have studied the optical and electronic properties of the MBE-grown non-doped, $5$ and $15\%$ $Co$-doped $ZnO$ thin films. SEM and AFM investigations have found that with growth of the $Co$ concentration the grain diameter is decreased, and one can observe spinodal component decomposition of $Zn_{1-x}Co_xO$ alloys. It results in inhomogeneous distribution of $Co$ atoms. There arise local regions enriched and depleted with $Co$ that possess the dimensions within the range $\sim15-20$ $nm$.

In the spectra of band-to-band PL, one can detect two bands, which is in accord with the possibility of creation of two regions with different the $Co$ concentration. The band that corresponds to a higher concentration of $Co$ atoms is shifted to the high-energy side as compared with the band corresponding to regions with a lower $Co$ concentration. The shift to the high-energy side is a result of $sp$-$d$ mixing of wave functions inherent to electrons of the conduction band with localized $d$-states of magnetic impurity.

In temperature-dependent PL spectra under excitation energies lying in the visible range, we have detected energy transitions in the $Co^{2+}$ center: $^2E(G)$, $^2A_1(G)$, $^2T_1(G)$, $^4T_1(P)$, $^2T_2(G)$ $\rightarrow$ $^4A_2(F)$, positions of which did not depend on the $Co$ concentration in $ZnO$ matrix.

Under band-to-band and sub-band excitation, we have detected resonant multi-phonon Raman spectra in $Zn_{1-x}Co_xO$ films. It has been investigated the dependence of the intensity for $LO$-phonon mode on the energy of exciting laser radiation. We have shown the possibility to study excited states of the $Co^{2+}$ center in $Zn_{1-x}Co_xO$ by using the resonant Raman effect. It allows studying the excited states of $Co^{2+}$ centers that have no luminescence.


\section*{References}
\begin{thebibliography}{10}
\bibitem{Thomas} Thomas D 1960 {\it J. Phys. Chem. Solids} {\bf 15} 86
\bibitem{Chambers} Chambers S 2010 {\it Adv. Materials} {\bf 22} 219
\bibitem{Schwartz} Schwartz D, Norberg N, Nguyen Q, Parker J, Gamelin D 2003 {\it J. Am. Chem. Soc.} {\bf 125} 13205
\bibitem{Ueda} Ueda K, Tabata H and Kawai T 2001 {\it Appl. Phys. Lett.} {\bf 79} 988
\bibitem{Dietl} Dietl T, Ohno H, Matsukura F, Cibert J, Ferrand D 2000 {\it Science} {\bf 287} 1019
\bibitem{Coey} Coey J, Venkatesan M, Fitzgerald C 2005 {\it Nat. Mater.} {\bf 4} 173
\bibitem{Kohan} Kohan A, Ceder G, Morgan D, Van de Walle C 2000 {\it Phys. Rev.} B {\bf 61} 15019
\bibitem{Dietl1} Dietl T, Spalek J 1982 {\it Phys. Rev. Lett.} {\bf 48} 355
\bibitem{Sato} Sato K, Katayama-Yoshida H, Dederichs P 2005 {\it Japanese J. of Appl. Phys.} {\bf 44} L948
\bibitem{Straumal} Straumal B, Mazilkin A, Protasova S, Myatiev A, Straumal P, Baretzky B 2008 {\it Acta Materialia} {\bf 56} 6246
\bibitem{White} White M, Ochsenbein S, and Gamelin D 2008  {\it Chem. Mater.} {\bf 20} 7107
\bibitem{Dietl2} Dietl T, Andrearczyk T, Lipinska A, Kiecana M, Tay M, Wu Y 2007 {\it Phys. Rev.} B  {\bf 76} 155312
\bibitem{Kuroda} Kuroda S, Nishizawa N, Takita K, Mitome M, Bando Y, Osuch K, Dietl T 2007 {\it Nature Materials} {\bf 6} 440
\bibitem{Yokoyama}  Yokoyama M, Yamaguchi H, Ogawa T, Tanaka M 2005 {\it J. Appl. Phys.} {\bf 97} 10D317
\bibitem{Samanta} Samanta K, Bhattacharya P, Katiyar R, Iwamoto W, Pagliuso P and Rettori C 2006 {\it Phys. Rev.} B {\bf 73} 245213
\bibitem{Hasuike} Hasuike N, Nishio K, Katoh H, Suzuki A, Isshiki T, Kisoda K, Harima H 2009 {\it J. Phys.: Condens. Matter} {\bf 21} 064215
\bibitem{Friedrich} Friedrich F, Nickel N 2007 {\it Appl. Phys. Lett.} {\bf 91} 111903
\bibitem{Guntherodt} G\"{u}ntherodt G and Zeyher R 1984 {\it Light Scattering in Solids IV, Springer series in Topics in Applied Physics} vol~54, (Berlin: Springer-Verlag) p~517
\bibitem{Chen} Chen Y, Liu  Y, Lu S, Xu C, Shao C, Wang C, Zhang J, Lu Y, Shen D, and Fan X, 2005 {\it J. Chem. Phys.} {\bf 123} 134701
\bibitem{Fonoberov} Fonoberov V, Alim K, Balandin A, Xiu F, and Liu J 2006 {\it Phys.Rev.} B {\bf 73} 165317
\bibitem{Lebedeva} Lebedeva N and Kuivalainen P 2002 {\it J. Phys.} C {\bf 14} 4491
\bibitem{Zeng} Zeng H, Cai W, Hu J, Duan G, Liu P and Li Y 2006 {\it Appl. Phys. Lett.} {\bf 88} 171910
\bibitem{Kurbanov} Kurbanov S, Panin G, Kim T, and Kang T 2009  {\it J. Lumin.} {\bf 129} 1099
\bibitem{Gong} Gong Y, Andelman T, Neumark G, O'Brien S, and Kuskovsky I 2007 {\it Nanoscale Res. Lett.} {\bf 2} 297
\bibitem{Zubiaga} Zubiaga A, Garc\'{\i}a J, Plazaola F, Tuomisto F, Saarinen K, Zu\~{n}iga P\'{e}rez J, and Mu\~{n}oz-Sanjos\'{e} V 2006 {\it J. Appl. Phys.} {\bf 99} 053516
\bibitem{Koidl} Koidl P 1977 {\it Phys. Rev.} B {\bf 15} 2493
\bibitem{Callender} Callender R, Sussman S, Selders M, and Chang R 1973 {\it Phys. Rev.} B {\bf 7} 3788
\bibitem{Cusco} Cusc\'{o} R, Alarc\'{o}n-Llad\'{o} E, Ib\'{a}\~{n}ez J, Art\'{u}s L, Jim\'{e}nez J, Wang B, and Callahan M 2007 {\it Phys. Rev.} B {\bf 75} 165202
\bibitem{Scepanovic} \v{S}\'{c}epanovi\'{c} M, Gruji\'{c}-Broj\u{c}in M, Sre\'{c}kovi\'{c} K 2010 {\it J. Raman Spectrosc.} {\bf 41} 914
\bibitem{Zigone} Zigone M, Vandevyver M, Talwar D 1981 {\it Phys. Rev.} B {\bf 24} 5763
\bibitem{Fonoberov1} Fonoberov F, Balandin A 2004 {\it Phys. Rev.} B {\bf 70} 233205
\bibitem{Serrano} Serrano J, Romero A, Manj\'{o}n F, Lauck R, Cardona M, and Rubio A 2004 {\it Phys. Rev.} B {\bf 69} 094306
\bibitem{Bergman} Bergman L, Chen X, Morrison J, Huso J, and Purdy A 2004 {\it J. Appl. Phys.} {\bf 96} 675
\bibitem{Kaschner} Kaschner A, Hoffmann A, and Thomsen  C 2001 {\it Phys. Rev.} B {\bf 64} 165314
\bibitem{Nielsen} Nielsen K, Bauer S, L\"{u}bbe M, Goennenwein S, Opel M, Simon J, Mader W, Gross R 2006 {\it phys. stat. sol.} (a) {\bf 203} 3581
\bibitem{Guntherodt1} G\"{u}ntherodt G, Merlin R, and Gr\"{u}nberg P 1979 {\it Phys. Rev.} B {\bf 20} 2834
\end{thebibliography}
\end{document}